\documentclass[conference]{IEEEtran}
\IEEEoverridecommandlockouts
\usepackage{cite}
\usepackage{amsmath,amssymb,amsfonts}
\usepackage{algorithmic}
\usepackage{graphicx}
\usepackage{textcomp}
\usepackage{xcolor}
\def\BibTeX{{\rm B\kern-.05em{\sc i\kern-.025em b}\kern-.08em
    T\kern-.1667em\lower.7ex\hbox{E}\kern-.125emX}}
\begin{document}

\title{Survey of Millimeter Wave Backscatter Communition Systems
\\
\thanks{Identify applicable funding agency here. If none, delete this.}
}

\author{\IEEEauthorblockN{1\textsuperscript{st} Weilin Chen}
\IEEEauthorblockA{\textit{University of Science and Technology of China} \\
\textit{}
Heifei, China \\
im26@mail.ustc.ecu.cn}

}

\maketitle

\begin{abstract}
Backscatter communication is a burgeoning low-power communication technology that has been introduced into the Internet of Things (IoT) due to its excellent self-sustainability. However, conventional backscatter communication (BackCom) technologies often suffer from insufficient data transmission rates, and are difficult to meet the increasing bandwidth demands of Internet of Things applications. In contrast, millimeter wave (mmWave) backscatter communication holds great potential for high-speed data transmission because of its rich spectrum resources. Therefore, this paper presents a comprehensive survey of the existing literature on mmWave backscatter communications. Firstly, we introduce the fundamental principles of mmWave backscatter communication, including its architecture, basic techniques and primer knowledge. Subsequently, we review the various mmWave backscatter communication techniques proposed in current literature. We then discuss some applications based on these techniques before finally addressing some challenges and open issues associated with mmWave backscatter.
\end{abstract}

\begin{IEEEkeywords}
backscatter communications, Internet of Things, millimeter wave
\end{IEEEkeywords}

\section{Introduction}
With the rapid development of IoT technology, there is an increasing demand for devices to be self-sustainable and have large-scale connectivity. In recent years, backscatter communication techniques have gained more attention due to advancements in backscatter modulation and energy harvesting techniques. The backscatter technology can significantly reduce the power consumption of a device by modulating its signal onto the signals transmitted by other devices, instead of generating and transmitting its own signal. For instance, current Bluetooth and WiFi-based backscatter techniques can significantly reduce the power consumption of backscatter transmitters (tags) by multiple orders of magnitude. The energy harvesting device integrated on the tag is capable of extracting energy from the incident signal to eliminate the need for batteries.

However, the current backscatter communication technology still suffers from a significant drawback: limited data transmission rate. The majority of backscatter technologies have a data transfer rate of less than 1Mb/s \cite{wu2022survey}. Although this speed may suffice for most IoT applications today, it remains inadequate for potential Internet applications such as augmented reality (AR). The primary cause of insufficient data transmission rates is the limited channel bandwidth, as exemplified by the 915MHz spectrum band commonly utilized in RFID technology \cite{FC2015}, which possesses a mere 500KHz channel bandwidth.

A feasible solution is to introduce millimeter wave technology. The unlicensed spectrum in the mmWave band has 200 times the bandwidth of the spectrum allocated to WIFI and RFID. This spectrum availability enables higher data transmission rates for mmWave backscatter communications. However, mmWave backscatter communication systems still face many challenges. A major limitation of mmWave technology is the significant signal attenuation caused by its high frequencies, necessitating beam alignment to concentrate energy in a specific direction \cite{hassanieh2018fast}. However, conventional beam search methods consume excessive power and are unsuitable for millimeter wave backscatter devices.

Although there exist survey articles on mmWave backscatter communication, these articles do not focus on millimeter wave backscatter communication systems. This paper is the first to provide a comprehensive overview on the state-of-the-art research on the architectures, system, and applications of mmWave BackComs. The contributions of this survey are summarized as follows: 

(1) This paper provides a discussion on fundamental principles of mmWave backscatter communication, including its architecture, basic techniques and primer knowledge

(2)This paper provides an exhaustive discussion on mmWave BackComs systems that have been proposed between 2019-2023. This paper provides a summary of different systems from the performance perspective.

(3)This paper provides a discussion on applications empowered by mmWave BackComs systems, and challenges that mmWave BackComs systems face.

The rest of this paper is organized as follows. Section \ref{sec2} presents the principles of mmWave BackComs including architecture, basic techniques, and primer knowledge. Section 3 provides a discussion on the current mmWave BackCom systems.  Section 4 describes applications demproved by mmWave BackComs. Challenges and open issues are discussed in Section 5. Finally, Section 6 concludes this paper. The abbreviations used in this article are summarized in Table \ref{tab1}.

\begin{table}[htbp] \label{tab1}
\caption{Abbreviations used and their meanings}
  \centering
    \begin{tabular}{|l|l|}
	\hline
	
    \textbf{Abbreviations} & \textbf{Meanings} \\
	\hline

\hline
	AP & Access Point \\
	\hline
	ASK & Amplitude Shift Keying \\
	\hline
    BackCom & Backcastter Communication \\
	\hline
	BER & Bit-error-rate\\
	\hline
	BPSK & Binary Phase Shift Keying \\
	\hline
	CDMA & Code Division Multiple Access\\
	\hline
	CW & Continuous Wave\\
	\hline
	FFT & Fast Fourier Transform\\
	\hline
	EH & Energy Harvesting\\
	\hline
	FM & Frequency Modulation\\
	\hline
	FMCW & Frequency Modulated Continuous Wave\\
	\hline
	FSK & Frequency Shift Keying \\
	\hline
	HAR & Human Activity Recognition \\
	\hline
	IF & Intermediate Frequency\\
	\hline
	IoT & Internet of Things\\
	\hline
	LoS & Line of Sight\\
	\hline
	MIMO & multiple-input multiple-output\\
	\hline
	mmWave & Millimeter Wave \\
	\hline
	NLoS & Non Line of Sight\\
	\hline
	OOK & on-off Keying\\
	\hline
	OFDM & Orthogonal Frequency Division Multiplexing\\
	\hline
	OTAM & Over The Air Modulation \\
	\hline
	TDMA & Time Division Multiple Access\\
	\hline
    \end{tabular}%
  \label{tab:addlabel}%
\end{table}%

\section{Principles of mmWave BackCom}\label{sec2}

In this section, we first present the architecture of a typical mmWave BackCom system. Subsequently, we introduce some fundamental techniques applied to mmWave BackCom systems. Finally, we discuss the primer knowledge related to mmWave.

\subsection{Architecture}

The typical components of mmWave BackComs systems include an ambient source, a backscatter tag, and a backscatter receiver \cite{boyer2013invited}. The backscatter tag transmits data to the backscatter receiver by reflecting the mmWave signal emitted from an ambient source, such as a mmWave radar. As shown in Fig.\ref{fig: arch}, the backscatter tag generally consists of an antenna module, an energy harvesting module, and a modulator module. The tag receives the incident signal through the antenna, collects energy from it through the energy harvesting module, and carries its own information on the modulated incident signal through the modulator module. Finally, the modulated signal is transmitted through the antenna.

\begin{figure}[!htp] 
\centering
\includegraphics[scale=0.24]{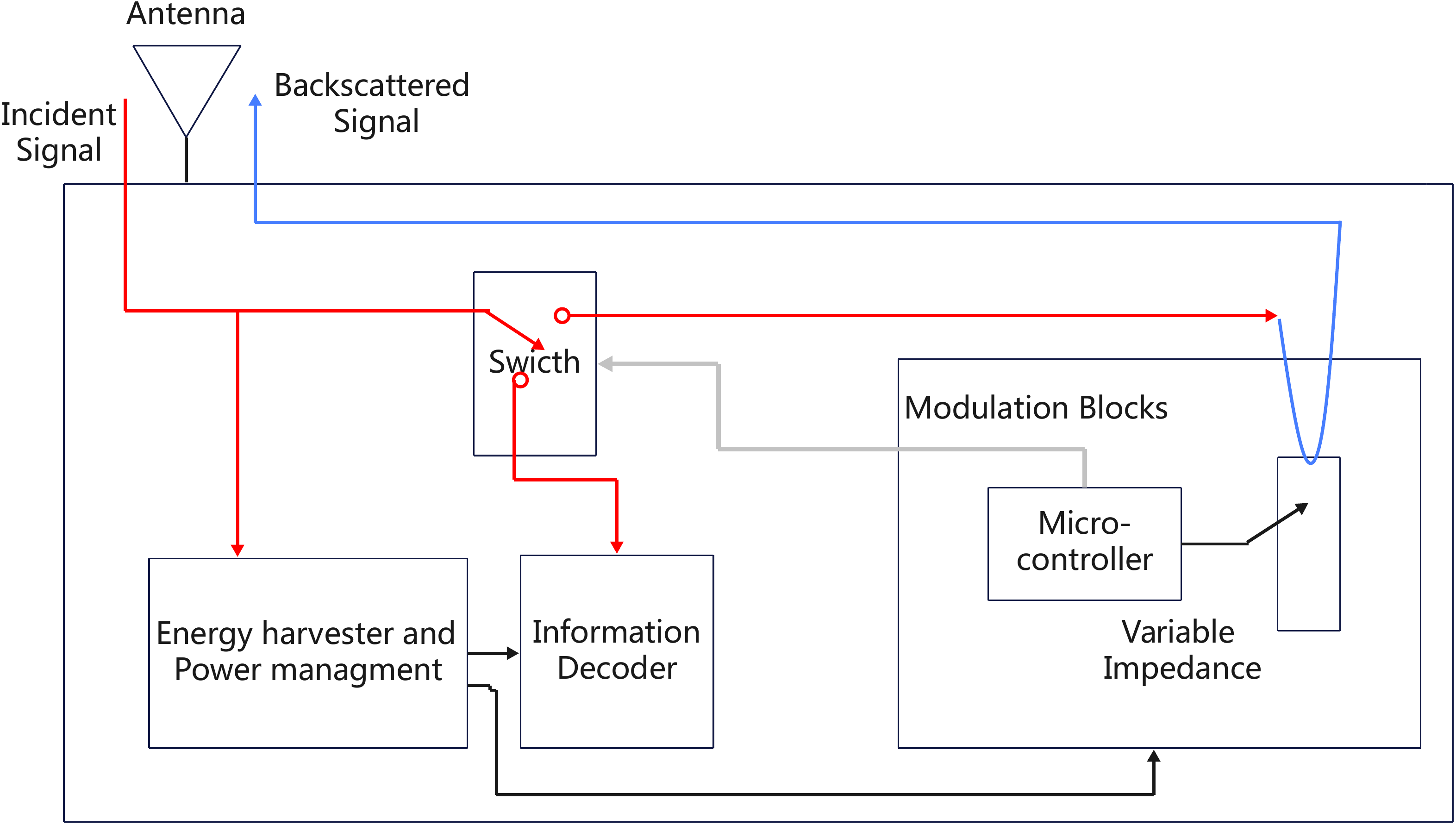}
\caption{The architecture of a backscatter tag.}
\label{fig: arch}
\end{figure}

\subsection{Basic techniques used in mmWave BackComs systems}

\subsubsection{Modulated backscatter}

The modulation of the incident signal by the backscatter tag is based on a characteristic of radio waves, namely that wireless signals are emitted at the junction of two media with different impedances, and the reflected signal will change depending on the difference between these media. Therefore, by adjusting the antenna load, modulation and reflection of the incident signal can be achieved. Moreover, the relationship between the backscattered signal and the incident signal can be calculated by Eq. (\ref{eq1}) \cite{xu2018practical, zhang2017freerider}:

\begin{equation}
\begin{aligned}
S_{out}&=S_{in}\times \Gamma_{tag}\label{eq1}\\
&= A_{in}e^{j(2\pi f_{in}t+\theta_{in})}\times \Gamma_{tag}
\end{aligned}
\end{equation}
where $S_{out}$ denotes the backscattered signal, $S_{in}$ denotes the incident signal and $\Gamma_{tag}$ denotes  the reflection coefficient of the antenna.

The reflection coefficient of the antenna can be computed by Eq. (\ref{eq2}):

\begin{equation}
\Gamma_{tag}=\frac{Z_{l}-Z_{a}}{Z_{l}+Z_{a}} = \left|\Gamma_{tag}\right|e^{j\theta_{tag}}\label{eq2}
\end{equation}
where $Z_a=\left|Z_a\right|e^{j\theta_{a}}$ denotes the antenna impedance and $Z_l=\left|Z_l\right|e^{j\theta_{l}}$ denotes the load impedance.

Combine Eq.(\ref{eq1}) and Eq.(\ref{eq2}), and we have:

\begin{equation}
\begin{aligned}
S_{out}&=A_{in}e^{j(2\pi f_{in}t+\theta_{in})}\times \Gamma_{tag}\label{eq3}\\
&= \left|\Gamma_{tag}\right|A_{in}e^{j(2\pi f_{in}t+\theta_{in}+\theta_{tag})}
\end{aligned}
\end{equation}

$\left|\Gamma_{tag}\right|$ and $\theta_{tag}$ can be computed as:

\begin{equation}
\left|\Gamma_{tag}\right|=\frac{\left|Z_{a}\right|^2+\left|Z_{l}\right|^2-\left|Z_{a}\right|\left|Z_{l}\right|cos(\theta_a-\theta_l)}{\left|Z_{a}\right|^2+\left|Z_{l}\right|^2+\left|Z_{a}\right|\left|Z_{l}\right|cos(\theta_a-\theta_l)}. \label{eq4}
\end{equation}

\begin{equation}
\theta_{tag}=arctan(\frac{2\left|Z_{a}\right|\left|Z_{l}\right|sin(\theta_a-\theta_l)}{\left|Z_{a}\right|^2-\left|Z_{l}\right|^2}). \label{eq5}
\end{equation}

We can infer that the backscattered signal will experience corresponding changes in amplitude and phase when altering the load impedance. Therefore, by switching between different loads, we can achieve amplitude modulation or phase modulation of the backscattered signal. In this way, the tag can modulate and piggy-back its own signal onto the incident signal. The operation of the backscatter antenna is easy to understand by assuming two states of reflection and non-reflection. OOK (On-Off Keying) is a modulation system \cite{tao2018ambient} that modulates the bits "0" and "1" of the backscatter transmitter onto the reflected signal. The transmission of data bit "1" indicates the reflection state of the backscatter antenna; Similarly, the transmission of data bit "0" switches the antenna to the no-reflection mode at the backscatter transmitter, the data sequence of "0" and "1" can be modulated into a reflected signal and then transmitted to the receiver. The receiver will then decode the data precisely based on the change in signal strength.

\subsubsection{Van Atta array}

The high power consumption and complexity inherent in traditional beam alignment solutions utilizing phased arrays and smart antennas render them unsuitable for BackCom tag design. This results in retrodirectivity, which reflects an incident signal solely towards the source direction. Van Atta array is a method for achieving retrodirectivity. As Fig.\ref{fig: arch} shows, Van Atta array consists of an array of antennas connected symmetrically in pairs by transmission lines of length equal to multiple of the guiding wavelength - that is, the transmission lines do not create additional phase differences for the incoming wave. Each antenna in this array serves as both a receiver and transmitter. The received signal from each antenna is transmitted through the line and reflected towards the source direction by its corresponding paired antenna. The Van Atta array can be designed entirely using passive components, making it an ideal choice for low power backscattering applications.

\begin{figure}[!htp] 
\centering
\includegraphics[scale=0.24]{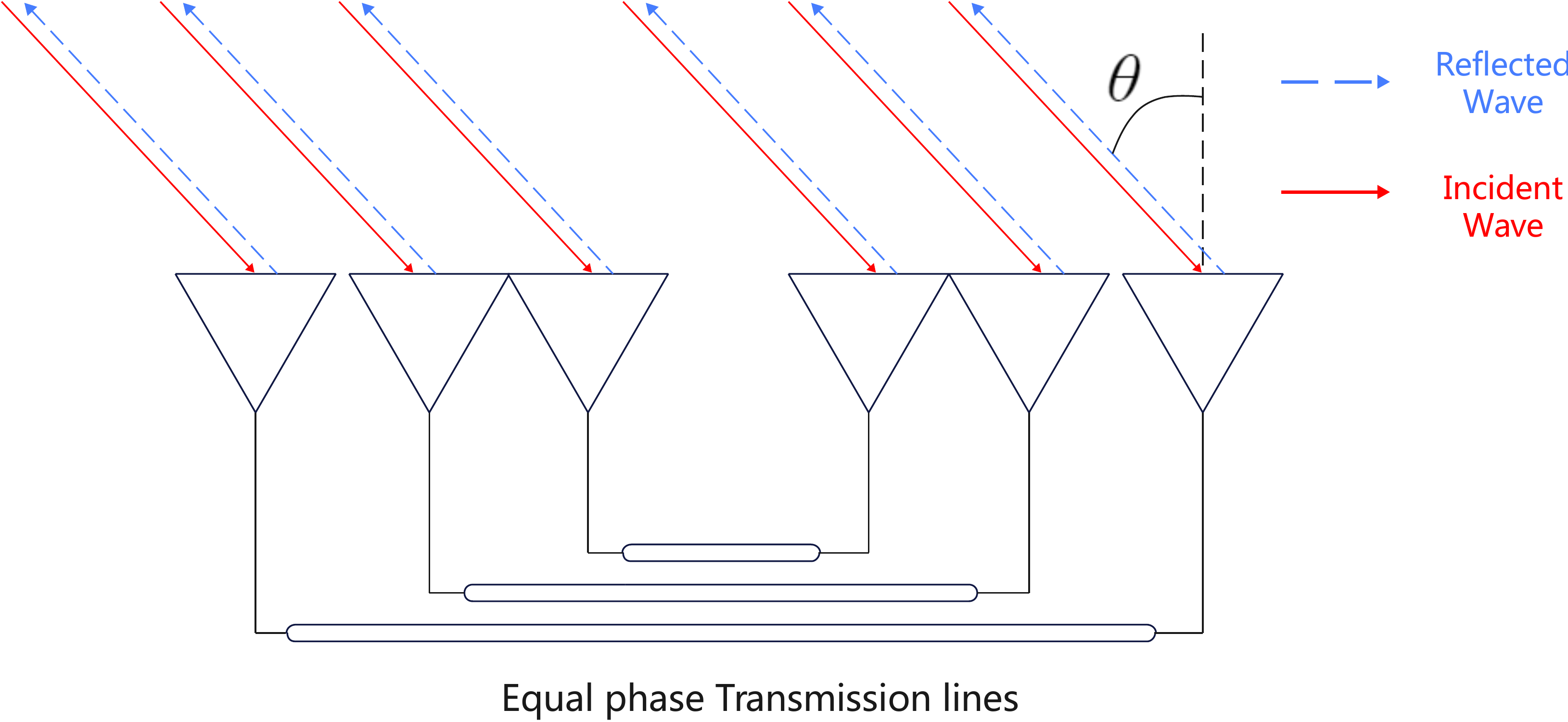}
\caption{The Van Atta antenna reflects the signal received by its mirrored antenna.}
\label{fig: arch}
\end{figure}

\subsubsection{Frequency Modulated Continuous Wave(FMCW) Rader}

FMCW radar refers to Continuous Wave (CW) radar whose transmit frequency is modulated by a specific signal. A signal transmitted by the FMCW radar is called a chirp, that is, a sinusoidal curve whose frequency varies with time. The signal modulation of FMCW radar mainly includes triangle wave modulation, sawtooth wave modulation, sine wave modulation and other different ways. The transmitted chirp is reflected back by the surrounding objects and returns to the radar with a propagation delay. The delayed transmitted and received chirps are mixed together, resulting in a single-tone IF signal whose power is concentrated at a frequency proportional to the time delay between them. The IF signal is subjected to FFT processing, yielding the target's range in relation to the radar. As such, the FMCW radar generates a distinctive frequency-range correlation, with the frequency of the signal being directly proportional to the distance of the target.By utilizing FMCW as the incident signal for mmWave backscatter, significant improvements can be achieved due to the ample mmWave bandwidth that facilitates chirp compression~\cite{barton1975radars}.

\subsection{Primer knowledge of mmWave}

With the rapid development of the communication industry, particularly in personal mobile communication, the low frequency radio spectrum has become saturated. Due to its wide bandwidth and short wavelength, millimeter wave technology can effectively address many challenges associated with high-speed wireless access, thus presenting a broad range of potential applications. Millimeter wave refers to an electromagnetic wave with a frequency range of 30GHz to 300GHz and a wavelength of 1 to 10mm. According to Shannon's theorem, the most direct and effective way to expand channel capacity is by increasing system bandwidth. Using the millimeter wave spectrum, which has a potential big bandwidth of 30~300 GHz, to provide higher data rates is considered a promising development trend for wireless communication technology in the future~\cite{rappaport2015millimeter}. However, despite its rich spectrum that enables high-speed wireless transmission, there are still some technical challenges that need to be addressed during the realization of millimeter wave communication~\cite{rangan2014millimeter}.

Due to the path loss is proportional to the square of the frequency, mmWave communications suffer from severe path losses. To mitigate this issue, large-scale antenna arrays and beamforming techniques are employed to focus energy in a specific direction for transmission~\cite{hassanieh2018fast}. Therefore, successful communication between two mmWave nodes is contingent upon the completion of beam alignment. In other words, prior to initiating communication, mmWave nodes must perform beam selection in order to determine the optimal transmission direction. To achieve this objective, most conventional schemes rely on phased array antennas for beam alignment~\cite{abari2016poster, mazaheri2019millimeter}. The phased array-based beam alignment scheme is not suitable for millimeter wave BackCom system tags. On the one hand, phased array antennas consume a significant amount of energy and are therefore unsuitable for low-power tags. On the other hand, phased array-based schemes require each node to have signal transmission and measurement capabilities, while mmWave BackCom tags can only reflect signals.

In recent years, all governments have allocated continuous unlicensed spectrum resources around 60GHz frequency. For example, the US classifies the unlicensed frequency range as 7GHz (57GHz-64GHz)\footnote{https://www.govinfo.gov/content/pkg/CFR-2007-title47-vol1/xml/CFR-2007-title47-vol1-sec15-255.xml}, Japan as 7GHz (59.4GHz-62.9GHz), and Europe as high as 9GHz (57GHz-66GHz). WirelessHD (WiHD) 1.0 standard is a wireless high-definition video transmission protocol running in the 60 GHz band~\cite{lawton2008wireless}. Its main purpose is to realize wireless high-speed data transmission between electronic devices within a few meters. The WiHD 1.1 standard can achieve 7 Gbit/s data transfer rate and 28 Gbit/s wireless transmission rate through spatial multiplexing. With 60 GHz as the carrier frequency band, IEEE 802.11 has formulated the IEEE 802.11ad standard for applications in hot spot coverage, office and other environments~\cite{ieee2012part}. IEEE 802.11ad uses array antennas to achieve array transmission coverage in WLAN, and the maximum transmission rate can reach 8 Gbit/s. The IEEE 802.11 organization also published the IEEE 802.11aj standard~\cite{ he2015complete}. Like other local area network standards published by IEEE 802.11, IEEE 802.11aj standard is mainly used to solve the wireless access of user terminals such as office Lans and shopping malls, and realize high-speed short-distance wireless coverage. IEEE 802.11aj adopts two transmission mechanisms, array antenna technology and multiple antenna technology, and realizes 15 Gbit/s wireless transmission with a coverage radius of 100 m by multiplexing technology (multiplexing up to 4 data streams)

\section{Mm-Wave BackCom systems}
This section discusses current mmWave BackCom systems. This section also summarizes and compares these systems in terms of advantages and disadvantages, methods, and performance.

\begin{figure}[!htp] 
\centering
\includegraphics[scale=0.24]{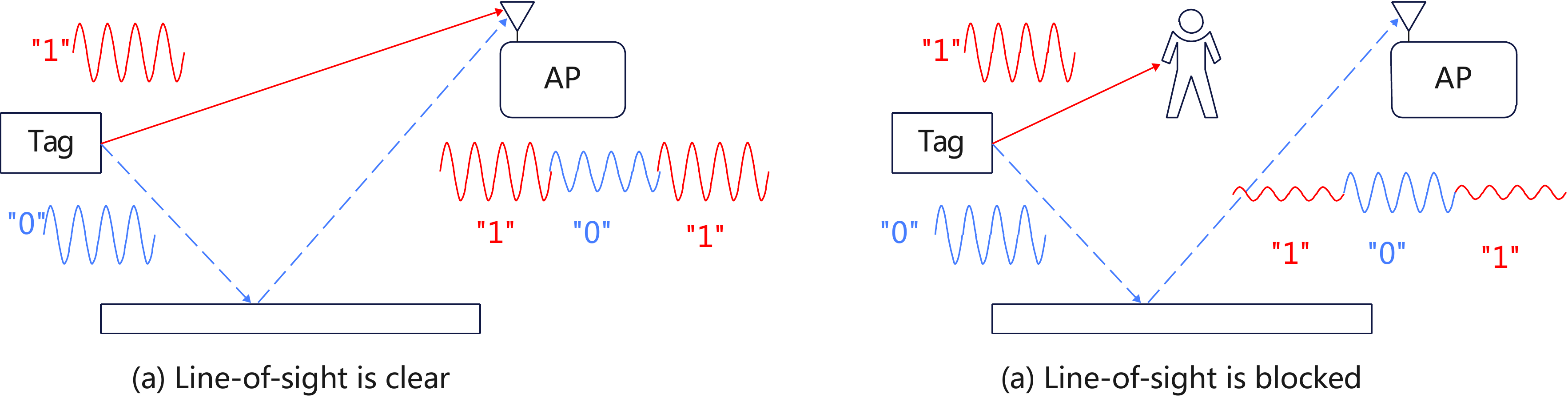}
\caption{Over The Air Modulation (OTAM) technique:  Instead of first creating an ASK signal and then choosing the best beam direction to transmit, OTAM sends a sine wave (carrier signal) to different beams depending on the value of data.}
\label{fig: mmX}
\end{figure}

Mazaheri et al.~\cite{mazaheri2019millimeter}introduce mmX, a mmWave BackCom sysytems based on the Over The Air Modulation (OTAM) technique.In the system, the tag reflects the backscattered signal in two directions, one is the LoS direction between the tag and the receiver, and the other is the NLoS direction. When the tag wants to modulate bit "1" onto the reflected signal, it will reflect the signal in the LoS direction. On the contrary, if the tag wants to carry bit "0" to the reflected signal, it will choose to reflect the signal in the NLoS direction.As Fig.\ref{fig: mmX} shows, there are two cases. When the LoS direction is not blocked, the amplitude of the reflected signal received in this direction is larger than that in the NLoS direction, and when the LoS direction is blocked, the amplitude of the reflected signal received in the LoS direction is smaller than that in the NLoS direction. Regardless of whether the LoS direction is blocked, the receiver can demodulate the reflected signal by the amplitude without the need to re-align the beam. Although mmX avoids re-alignment of beams in case of blockage in the LoS direction through the OTAM technique, it still has some serious shortcomings. As previously mentioned, the tag node lacks beam search capabilities. As a result, the direction of the reflected signal from the tag in mmX system is fixed. This also means that any changes to the relative position between the tag and receiver will prevent reception of the reflected signal by the receiver. These limitations greatly restricts applications for mmX technology.The experimental results show that mmX can achieve a throughput of 100 Mbps at 18 m, which is higher than current BackCom based on BLE, WiFi and other signal source.

Mazaheri et al.~\cite{mazaheri2021mmtag}then introduce mmTag, a mmWave BackCom sysytems utilizing the Van Atta array for passive beam searching. In passive beam searching, AP complete beam searching to find tag nodes while tags only reflect the AP’s signal back in the direction of arrival using Van Atta array. Passive beam search exploits the properties of millimeter waves. The transmission mode of millimeter wave is close to that of visible light, which has a reversible optical path. Therefore, the transmission path of millimeter wave is also reversible, as long as the incident signal is reflected back along the original way of incidence angle, the reflected signal will certainly reach the signal source. In this way, mmTag leaves the task of beam search entirely to the AP and avoids beam alignment by exploiting the characteristics of Van Atta antennas at the tag. MmTag also uses OOK to modulate the signal, and polarization conversion is used to avoid interference between reflected and incident signals. ompared to mmX, mmaTag's passive beam search does not require prior tag placement and allows for arbitrary positioning between the tag and AP, thus increasing its practicality. The experimental results show that mmTag can achieve a throughput of 100 Mbps at 8 m, with 2.4 nJ/bit of power consumption.

Although mmTag successfully solves the beam alignment problem in millimeter wave backscatter communication, it still has some obvious shortcomings. For example, it does not consider making full use of mature commercial components and technologies, and does not consider the performance of the system in outdoor scenarios, mobile scenarios and NLoS scenarios. Bae et al.~\cite{bae2022omniscatter} propose omniscatter, which takes full advantage of mature commercial FMCW radars and supports complex scenarios such as outdoor, mobile and NLoS. Omnisccatter proposed HD-FMCW based on FMCW. By improving symbol, HD-FMCW can separate the noise (environmental reflection) from the tag signal and isolate it into a special bin while retaining the original advantages of FMCW. This allows omniscatter to achieve extreme sensitivity and robust communication regardless of the environment and scenario. The Omniscatter switches impedance at the tag to FSK the incident signal and demodulates the reflected signal at the receiver using the FFT operation of the FMCW radar. Omniscatter also presents a coordination-free FDMA. Tags are passively allocated with the corresponding channels by their tag-radar distances because of the property of FMCW. Experiments show that compared with the previous system, omniscatter performs better in complex scenarios, which can be applied to outdoor, mobile, NLoS and other scenarios, and can realize BackCom over a longer distance (40m).

An et al.~\cite{an2022millimeter} also presents a mmWave BackCom systems based on FMCW rader. In this system, the FMCW radar on the vehicle acts as a signal source to transmit the incident signal, and the tag reflects the signal to the FMCW radar through the van atta array after receiving and modulating the signal. The collision between multiple tags is avoided by using CDMA. For each tag, a separate signature code is used to modulate the raw data of the baseband. On the radar side, the received signal will be associated with each signature code to recover the data for each tag. In this case, correlation increases the signal-to-interference ratio (SIR) and signal-to-noise ratio (SNR) of the target tag signal. Thus, the data from each label can be distinguished. Not only that, the normal radar signal can guarantee its SIR To realize sensing while have the ability to recover data from the tag. In other words, this system can achieve millimeter wave backscatter communication while ensuring the basic functions of FMCW radar. The experimental results show that mmTag can achieve a throughput of 8 Gbps at 0.65 m. However, the system exhibits certain deficiencies as the authors have not provided a corresponding code allocation scheme. Given the mobile nature of vehicle radar, this lack of an appropriate code allocation scheme renders the system infeasible.

\textbf{Discussion: }In this section, we have reviewed mmWave BackCom systems. Table \ref{tab:str},\ref{tab:tech}and \ref{tab:perf} give a summary about the advantages and disadvantages, techniques and performance of these systems, respectively. According to the table, we observe that omniscatter~\cite{bae2022omniscatter} has the best performance. Compared with mmTag~\cite{mazaheri2021mmtag}, omniscatter is more suitable for complex scenarios and takes full advantage of mature commercial components. Compared with the scheme proposed by An et al.~\cite{an2022millimeter}, omniscatter provides coordinate-free FDMA and has better feasibility. Despite this, omniscatter still has some problems, such as the bit error rate is still high in complex environments, and the communication distance of backscatter is still insufficient. Therefore, there is a need for further studies to address these problems.

\begin{table}[htbp] 
\caption{Strengths and identified gaps in existing studies}
  \centering
    \begin{tabular}{|l|l|l|l|}

	\hline
	
    \textbf{Systems} & \textbf{Year} & \textbf{Strenghth(s)} & \textbf{Gap(s) in studies}\\
	\hline
\hline
	&  & OTAM is used to solve the  & The receiver has to  
\\
\cite{mazaheri2019millimeter} &2019&challenge of beam searching &be fixed to receive\\
&&and NLoS.& the signal of the tag. \\
	\hline
	 &  &The dedign of passive beam  & The scenerios of   
\\
\cite{mazaheri2021mmtag}&2021&searching reflect the signal& mobility, multi-tags\\
&&to the  source. &and NLoS are\\
&&&not considered. \\
	\hline
     &  &HD-FMCW is proposed to   &The BRE is high\\
\cite{bae2022omniscatter} &2022&enable distance-based &in complex\\
&&automatic channel allocation &environments.  \\
&&and lightweight modulation.&\\
	\hline
	 &  &CDMA with FMCW are   &Lack of  code \\
\cite{an2022millimeter}&2022&combined to avoid collisions& allocation schemes.\\
&&between multi-tags.&\\
	\hline
    \end{tabular}%
  \label{tab:str}%
\end{table}%

\begin{table}[htbp] 
\caption{Summary of techniques used in mmWave BackCom systems}
  \centering
    \begin{tabular}{|l|l|l|l|}
	\hline
	
    \textbf{Systems} & \textbf{Data} & \textbf{Beam} & \textbf{Self-Interference}\\
	&\textbf{Modulation}&\textbf{Alignment}&\\
	\hline
\hline
	\cite{mazaheri2019millimeter} & OOK  & OTAM   & Null\\

	\hline
	\cite{mazaheri2021mmtag} & OOK  &Van Atta array  & Polarization Conversion \\
	\hline
    \cite{bae2022omniscatter}  & FSK &Van Atta array &HD-FMCW\\
	\hline
	\cite{an2022millimeter} &BPSK  & Van Atta array   &FMCW+CDMA \\
	\hline
    \end{tabular}%
  \label{tab:tech}%
\end{table}%

\begin{table}[htbp] 
\caption{Summary of preformance of mmWave BackCom systems}
  \centering
\resizebox{\columnwidth}{!}{
    \begin{tabular}{|l|l|l|l|l|l|}
	\hline
	
    \textbf{Systems} & \textbf{Throughput}  & \textbf{Range}&\textbf{BER}&\textbf{Mobility}&\textbf{Multiplexing}\\
	\hline
\hline
	\cite{mazaheri2019millimeter} & 100Mbps     & 18m&$10^{-3}$&No&No\\
	 & (at 18m)  &   &(at 8m)&&\\
	\hline
	\cite{mazaheri2021mmtag}  & 100Mbps     & 14m&$10^{-3}$&No&No\\
	 & (at 8m)  &   &(at 8m)&&\\
	\hline
    \cite{bae2022omniscatter}   &     & 40m&$<10\%$&Yes&Yes\\
	 & &   &(at 40m)&&\\
	\hline
	  & 8Gbps(at     & 18m&$1.93\times$&Yes&Yes\\
	\cite{an2022millimeter}&  0.65m)  &   &$10^{-5}$(at &&\\
	 &   &   & 0.65m)&&\\
	\hline
    \end{tabular}%
}
  \label{tab:perf}%
\end{table}%

\section{Applications}

Millimeter wave backscatter communication has the characteristics of low energy consumption and high data rate, so it has a good application prospect in many fields (e.g., localization~\cite{bae2023hawkeye,soltanaghaei2021millimetro}, logistics\cite{adeyeye2019miniaturized}, additive manufacturing and inkjet printing~\cite{kimionis2021printed}). In this section, we provide a brief discussion on the applications based on mmWave BackComs.

\subsection{Localization}
Soltanaghaei et al.~\cite{soltanaghaei2021millimetro} introduces Millimetro, an ultra-low-power tag capable of high-precision localization over long distances. The tag is designed for locating roadside infrastructure such as lane markings and road signs in the context of autonomous driving. While RF-based localization offers a natural solution, current ultra-low-power systems face challenges in achieving accurate operation at extended ranges under stringent latency requirements. Millimetro overcomes this challenge by utilizing existing millimeter-wave FMCW radars operating at a frequency with sufficient bandwidth to ensure high precision and minimal latency.

Bae et al.~\cite{bae2023hawkeye} introduces Hawkeye, a novel mWwave backscatter technology that enables simultaneous 3D localization over hundreds of scales and distances up to hundreds of meters with subcentimeter accuracy for the first time. The Van Atta array design of the Hawkeye tag features reverse reflectivity in both pitch angle and azimuth plane, which effectively suppresses multipath interference while providing precise 3D localization. The Hawkeye localization algorithm is a lightweight signal processing method that is compatible with commercial FMCW radars. It uniquely exploits the interaction between tag signals and clutter, utilizing spectral leakage for fine-grained localization, thereby enabling accurate localization of numerous objects in a vast area.

\subsection{Logistics}

In this paper \cite{adeyeye2022machine}, a miniaturized mmWave RFID tag with unique identification and spatial localization capabilities is proposed for the first time. The FMCW radar is used as the reader to realize the spatial localization of RFID tags. This radar is used to resolve the modulated backscattered signal returned by the RFID tag. Spatial information -distance, Angle with respect to the reader antenna -is contained in the peak frequency and phase of the returned signal. The millimeter wave compatibility of the RFID tags presented in this paper makes them suitable for various emerging 5G and IoT topologies and architectures that require identification and/or localization. The device has an extremely low form factor and ultra-low power consumption, making it well suited for various low-power sensing and detection applications. Moreover, for dense implementations, the low bandwidth occupation (100s of Hz) of each tag, suggests that a large number of frequency-separated tags can be uniquely identified and spatially localized.

\subsection{Additive manufacturing and inkjet printing}

Since the size of the antenna is generally proportional to the wavelength of the corresponding radio wave, antennas in mmWave communication systems are typically small. This characteristic results in very low form factor components for millimeter wave communication, which enables its application across a wide range of fields. This paper presents a mmWave modulator and antenna array designed for gigabit data-rate backscatter communications~\cite{kimionis2021printed}. The RF front-end comprises a microstrip patch antenna array and a single pseudocrystal high electron mobility transistor that supports multiple modulation formats, including binary phase shift keying, quadrature phase shift keying, and quadrature amplitude modulation. The circuit was fabricated using nano-silver particle ink on a flexible liquid crystal polymer substrate via inkjet printing technology.

\section{Open issues and challenges}
In this section, we provide a discussion of some remaining open issues and challenges in the area of mmWave BackComs.

\subsection{Beam alignment}

Although the Retro-directive based on Van Atta array has been used to solve the beam alignment challenge in current millimeter-wave backscatter communication systems \cite{mazaheri2021mmtag, bae2022omniscatter, an2022millimeter}, there are still some shortcomings in this solution. The Retro-directive scheme based on the Van Atta array can only return the incident signal the same way, that is, the tag can only return its own information to the AP that actively sends signals to him. This significantly limits the application scenarios of mmWave backscatter communication. Therefore, it is necessary to further research into more active beam alignment scheme to solve this problem. To solve this problem, Janjua et al.~\cite{janjua2023beam} proposed a solution based on symbiotic radio. The beam selection method based on amplitude maximization is implemented with the help of symbiotic radio, so as to improve the rate and performance of the system.

\subsection{Multiple-access}

In a multiple-access BackCom system, the primary challenge lies in effectively managing collisions that arise from concurrent tag transmissions. To address this issue, conventional MAC schemes such as space/frequency/code/time division multiple access can be employed to prevent collisions during multi-tag transmissions. The Omniscatter \cite{bae2022omniscatter} and the system proposed in \cite{an2022millimeter} utilize frequency division multiple access and code division multiple access respectively to enhance scalability. The challenge that arises is the coordination of tags. Due to their inherent limitations, communication and coordination among tags become arduous tasks. Therefore, a multi-access scheme with an appropriate coordination mechanism is imperative for mmWave BackCom systems.

\subsection{Security}

Security in MmWave BackCom is a novel topic. This paper presents a potential attack scenario in millimeter-wave backscatter communication \cite{lazaro2022spoofing}. The authors utilize backscatter tags to inject false information into the received signal of FMCW radar, and propose a spoofing attack detection method based on the change of frequency scan slope between frames. This paper demonstrates a proof-of-concept spoofing device for the 24 GHz ISM band. The spoofing device is based on an amplifier that connects two antennas and modulates the switching amplifier bias. The use of amplifiers enables us to increase the level of spoofing signal compared to other modulated backscatter methods. Simulation and experimental results demonstrate that this proposed method can generate a pair of fake targets at different distances and speeds.

\subsection{Health care}

Due to the unique wavelength characteristics of millimeter waves, the antenna size required for mmWave BackCom is generally very small. As a result, compared with other BackCom systems, mmWave BackCom systems have significant advantages in the field of wearable devices. This opens up more possibilities for mmWave BackCom applications, such as healthcare. By integrating mmWave BackCom into wearable devices, we can provide users with more transparent and ubiquitous healthcare services.

\section{Conclusion}

Considering the increasing number of Internet of Things applications and the increasingly crowded low spectrum band, the abundant spectrum resources owned by millimeter wave backscatter communication system have great application value. In this article, we have provided a detailed discussion of all mmWave BackCom systems published between 2019 and 2023. We summarize and compare these systems, discussing their advantages and disadvantages, techniques, and performance. Subsequently we introduce and discuss the application based on mmWave BackCom. We also discuss challenges and open issues regarding mmWave BackCom.

\bibliographystyle{plain}
\bibliography{ref}

\vspace{12pt}

\end{document}